\documentclass[sigplan,10pt]{acmart}
\AtBeginDocument{%
  \providecommand\BibTeX{{%
    \normalfont B\kern-0.5em{\scshape i\kern-0.25em b}\kern-0.8em\TeX}}}

\setcopyright{acmlicensed}
\copyrightyear{2024}
\acmYear{2024}
\acmDOI{XXXXXXX.XXXXXXX}

\acmBooktitle{4th Workshop on Machine Learning and Systems (EuroMLSys ’24), April 22, 2024, Athens, Greece}
\acmISBN{978-1-4503-XXXX-X/18/06}

\usepackage{xspace}
\begin{document}

\newcommand\eg{\emph{e.g.},\xspace}
\newcommand\ie{\emph{i.e.},\xspace}
\newcommand\etc{\emph{etc}.\xspace}
\newcommand\via{\emph{via}}
\providecommand{\etal}{\emph{et al.}\xspace}

\newcommand{\para}[1]{\smallskip \noindent \textbf{#1}}
\newcommand{\parait}[1]{\smallskip \noindent \textit{#1}}

\newcommand{\name}{\textsc{GuaranTEE}\xspace}

\newcommand{\numberedcircle}[3]{% 
    \begin{tikzpicture}
        % Draw the circle
        \draw[color=black, fill=white] (0,0) circle (#1);
        % Place the number in the center of the circle
        \node at (0,0) {\textcolor{black}{#2}};
    \end{tikzpicture}%
}
\title{\name: Towards Attestable and Private ML with CCA}

% \author{}
% \affiliation{%
%   \institution{}
%   \streetaddress{}
%   \city{}
%   \country{}}
% \email{}

\author{Sandra Siby, Sina Abdollahi, Mohammad Maheri, Marios Kogias, Hamed Haddadi}
\affiliation{
  \institution{Imperial College London}
  \country{}
}

\if 0
\author{Lars Th{\o}rv{\"a}ld}
\affiliation{%
  \institution{The Th{\o}rv{\"a}ld Group}
  \streetaddress{1 Th{\o}rv{\"a}ld Circle}
  \city{Hekla}
  \country{Iceland}}
\email{larst@affiliation.org}

\author{Valerie B\'eranger}
\affiliation{%
  \institution{Inria Paris-Rocquencourt}
  \city{Rocquencourt}
  \country{France}
}

\author{Aparna Patel}
\affiliation{%
 \institution{Rajiv Gandhi University}
 \streetaddress{Rono-Hills}
 \city{Doimukh}
 \state{Arunachal Pradesh}
 \country{India}}

\fi 

% \affiliation{%

% \affiliation{%

% \affiliation{%

% \affiliation{%

\renewcommand{\shortauthors}{}

\begin{abstract}
Machine-learning (ML) models are increasingly being deployed on edge devices to provide a variety of services.
However, their deployment is accompanied by challenges in model privacy and auditability.
Model providers want to ensure that (i) their proprietary models are not exposed to third parties; and (ii) be able to get attestations that their genuine models are operating on edge devices in accordance with the service agreement with the user. 
Existing measures to address these challenges have been hindered by issues such as high overheads and limited capability (processing/secure memory) on edge devices.

In this work, we propose \name, a framework to provide attestable private machine learning on the edge.
\name uses Confidential Computing Architecture (CCA), Arm's latest architectural extension that allows for the creation and deployment of dynamic Trusted Execution Environments (TEEs) within which models can be executed.
We evaluate CCA's feasibility to deploy ML models by developing, evaluating, and openly releasing a prototype. 
We also suggest improvements to CCA to facilitate its use in protecting the entire ML deployment pipeline on edge devices.
\end{abstract}





\maketitle

\section{Introduction}
Machine-learning models are increasingly being deployed on edge devices (such as smartphones, IoT gateways, and home routers) for various purposes such as health monitoring, anomaly detection, face recognition, voice assistants, handwriting recognition \etc
Running local models on edge devices provides several advantages over cloud-based approaches. 
Sensitive user data from the devices do not have to be sent to external model providers for inference, thereby providing privacy benefits~\cite{10.1145/2185376.2185390}. 
Running models locally avoids the need for large data transfers, which can be costly in terms of latency and bandwidth. 
Finally, local models facilitate private personalization based on user preferences~\cite{8366985, sang2023beyond}. 

At the same time, models on edge devices pose challenges.
Model providers are increasingly demanding \textit{model privacy} and protection, \ie to ensure that their proprietary model information (\eg weights) are not exposed to external parties. 
Providers also desire \textit{model verifiability and attestability}, \ie to ensure that their models have run on the device as expected and have not been tampered with.

Prior work has shown that on-device models, such as those on mobile phones, are susceptible to various attacks~\cite{xu2019first, sun2021mind}.
Sun \etal~\cite{sun2021mind} showed that deep-learning models on mobile phones have insufficient protections -- their analysis of $\approx$ 40,000 apps on the Android store revealed that 41\% of apps lack any form of protection, and in cases where some protection such as encryption exist, they can be overcome by simple attacks.

Various solutions have been proposed for model protection.
Watermarking techniques can detect model theft but not prevent them, and are susceptible to tampering~\cite{boenisch2021systematic, xue2021intellectual}.
Cryptographic techniques such as homomorphic encryption (HE)~\cite{orlandi2007oblivious, gilad2016cryptonets, van2019sealion} or secure multiparty communication (SMC)~\cite{mohassel2017secureml, riazi2018chameleon} are hindered by computational and communication overheads.
Hardware-assisted techniques, using trusted execution environments (TEEs), are a performant alternative to cryptographic techniques. 
However, many TEE solutions are tailored towards cloud environments (\eg using Intel SGX) and not applicable to edge devices which have limited memory and computational power. 
Hardware-assisted solutions on edge devices have primarily focused on how models can be deployed on edge devices with limited TEE memory.
Several solutions involve either partitioning models and running part of the model within TEEs~\cite{mo2020darknetz, hou2021model, shen2022soter, sun2021mind}, or pruning models before deployment so that they can fit within TEE memory~\cite{hu2023secure}. 
The few works that have experimented with deploying entire models within TEEs are limited by the number of enclaves they can run in parallel and lack of support for secure peripherals~\cite{bayerl2020offline}.

In this work, we propose \name, a framework to deploy and run machine-learning models on the edge in a private and verifiable manner. 
\name is motivated by the introduction of Arm CCA (Confidential Computing Architecture)~\cite{ccasite, cca} -- a set of extensions in Arm's new architecture that allow for the creation of dynamic, hardware-protected enclaves, \textit{realms}.
TrustZone(security extensions in Arm's previous architecture)~\cite{trustzone} is widely deployed on edge devices, but is not appropriate for running ML models due to security and memory limitations~\cite{brasser2019sanctuary}.
CCA's features, which are tailored towards deploying ML at the edge, and its presence in Arm's next architecture make it a promising candidate for widespread deployment on edge devices. 
Thus, in this work, we explore CCA's potential and limitations to implement \name.

Our contributions are as follows:

\begin{itemize}
    \item We develop \name, a framework that allows for machine-learning models from a provider to be run on end devices in a private and verifiable manner. We explore Arm's new Confidential Computing Architecture (CCA) to implement \name. In \name, a model runs within a CCA \textit{realm} -- a hardware-protected enclave that can be established at runtime. Using CCA allows for entire ML models to be run within a trusted and private environment, without resorting to partitioning. 
    \item We develop and evaluate a prototype of \name using Arm's Fixed Virtual Platforms (FVP) simulator~\cite{fvp}\footnote{Our code is available at: \url{https://github.com/comet-cc/GuaranTEE}}. Our preliminary results indicate that running a TensorFlow Lite image-recognition model within a realm for inference results in an overhead of 1.7 times the number of instructions required to run it within a normal world virtual machine. 
    \item We discuss challenges involved in implementing our prototype using CCA, and potential enhancements to the CCA architecture to enable better protection of the ML pipeline on edge devices. 
\end{itemize}

%Add challenges section

%Guarantee architecture: with CCA coming, how can we use that 

%preliminary results 

%future work 

\section{Model protection on the edge}
Running machine-learning models on edge devices involves local storage of models and hence, poses several challenges with regard to \textit{model privacy} and \textit{model verifiability}.
Model providers desire model privacy as their models might be proprietary; providers do not want to expose any information about their models to external parties such as other (potentially competing) model providers, the end user, or malicious actors. 
Model providers also require model verifiability, to ensure that their models exhibit the expected behavior and have not been tampered with. 
Leaked or modified models, especially in security-critical applications such as banking or medical services, can result in harm to the end user \eg identity theft or exposure of personal information~\cite{sun2021mind}.

Prior work has shown that on-device models, such as those on mobile phones, are susceptible to model stealing and adversarial machine learning~\cite{deng2022understanding, huang2022smart, hu2023first, sang2023beyond} and lack necessary protections~\cite{xu2019first, sun2021mind}.
%
%For example, Sun \etal~\cite{sun2021mind}'s analysis of $\approx$ 40,000 apps on the Android store revealed that 41\% of apps lack any form of protection, and even in cases where some protections such as encryption exist, they can be overcome by simple attacks.
%
Thus, it is necessary to develop techniques for model protection that can provide better privacy and verifiability.

There are several techniques aimed at providing model protection.
Techniques such as watermarking and fingerprinting allow model providers to identify stolen models~\cite{li2021survey, liu2023provenance, sun2023deep}.
However, they are \textit{passive defenses}, \ie they do not prevent theft but detect it afterward~\cite{boenisch2021systematic}. 
In addition, works have successfully demonstrated evasion, removal, and tampering attacks against these techniques~\cite{xue2021intellectual}.
Another category of techniques involves \textit{cryptographic} means to provide private and secure machine learning.
These primarily include: homomorphic encryption (HE), which allows for operations to be run directly on encrypted data (\eg~\cite{orlandi2007oblivious, gilad2016cryptonets, van2019sealion}), and secure multi-party computation (SMPC), where multiple parties jointly compute a function over inputs that are kept private (\eg~\cite{mohassel2017secureml, riazi2018chameleon}), or a combination of the two (\eg~\cite{liu2017oblivious}).
These techniques have even been applied in edge scenarios(\eg~\cite{lin2022performance}).
However, cryptographic techniques are hindered by high computation (for HE) and communication (for SMPC) overheads.

\textit{Hardware-assisted techniques} address the performance limitations of cryptographic techniques.
They make use of Trusted Execution Environments (TEEs), which are isolated processing environments in which applications can be executed securely~\cite{mo2022sok}.
There are several works that use TEEs for private inference in cloud-based applications (such as MLaaS platforms), mainly using Intel SGX~\cite{tramer2018slalom, gu2018yerbabuena, hashemi2020darknight}.
However, these solutions are not appropriate for edge devices which have limited memory and computational power.
Hardware-assisted techniques on the edge focus on how machine-learning models can be deployed in devices with limited capabilities.
Several works have proposed putting a subset of a model within the TEE and offloading the rest to untrusted accelerators -- these include shielding deep layers~\cite{mo2020darknetz}, shallow layers~\cite{hou2021model}, intermediate layers~\cite{shen2022soter}, non-linear layers~\cite{sun2023shadownet} within a TEE. 
Zhang \etal~\cite{zhang2024tsdp} termed these solutions as \textit{TSDP} or TEE-Shielded DNN Partition.
Zhang \etal showed that TSDP solutions are vulnerable to privacy attacks.
Both Zhang \etal and Liu \etal~\cite{liu2023mirrornet} showed that TSDP solutions are vulnerable to privacy attacks.
They proposed two methods of protecting models while accounting for the memory limitations of TEE.
Zhang \etal proposed TEESlice, a paritioning-before-training approach which partitions a model into a lightweight private part which resides within the TEE, and a public backbone outside the TEE.
Liu \etal proposed MirrorNet, where a model is trained using a combination of a backbone model outside the TEE and a lightweight network within the TEE.
Both these approaches involve separating the contents of the model such that confidential information is protected within the TEE. 

Due to TEE memory limitations, there has been limited work that has explored deploying an entire model within a TEE.
Brasser \etal~\cite{brasser2019sanctuary} designed Sanctuary, an architecture that allows for the creation of isolated user-space enclaves in the normal world on top of TrustZone. 
Bayerl \etal's~\cite{bayerl2020offline} work on Offline Model Guard (OMG) used Sanctuary to run machine-learning models in user-space enclaves in edge devices.
However, Sanctuary's core-based protection limits the number of concurrent isolated enclaves to the number of cores and does not support secure peripheral access~\cite{sun2022leap}.
Sun \etal~\cite{sun2022leap} proposed LEAP to overcome these limitations. 
Both OMG and LEAP rely on TrustZone, which has limitations that we discuss in Section~\ref{subsec:cca}.
Hu \etal proposed an orthogonal pruning approach to reduce the size of the model and enable it to be run within a TEE~\cite{hu2023secure}. 

In this work, we revisit the possibility of running an entire model within the TEE.
Our work is motivated by the introduction of Arm's Confidential Computing Architecture (CCA) which allows for the dynamic creation of hardware-protected enclaves called realms.
CCA is the next version of Arm's architecture to enable secure execution environments, in parallel with TrustZone.
As TrustZone is already widely deployed on end devices, we envision that CCA will also see real-world deployment in the near future.
In the next section, we describe CCA in more detail.

\subsection{Towards CCA}
\label{subsec:cca}

In this section, we describe TEE support on Arm.
We first provide an overview of TrustZone and its limitations in running machine-learning models before delving into CCA.
We also briefly describe existing work on CCA.

\para{Arm TrustZone.} TrustZone refers to hardware security extensions that were introduced by Arm in their Armv6K architecture in 2004 and enables the creation of TEEs~\cite{trustzone, pinto2019demystifying}.
TrustZone allows for the creation of two execution environments (or \textit{worlds}) that are isolated from one another: the Normal World and the Secure World.
The processor can be in one of these worlds at any point in time.
The transition between the worlds is managed by the highest-privileged firmware in the system, called the Secure Monitor.
%\footnote{Secure monitor should change a bit in Secure configuration register SCR\_EL3.NS. This bit is always checked during CPU accesses to memory.}. 
%
Each world has its own dedicated memory -- When the processor is in the non-secure state (\ie operating in the normal world), software in the normal world cannot access the secure world's memory. 
There is no restriction on access to the normal world memory when the execution is given to the secure world. 
The normal world usually runs a rich software stack which can include an operating system, applications, hypervisor, \etc whereas the secure world runs a smaller stack which includes a lightweight kernel (Trusted OS) supporting several security-sensitive services (Trusted apps) such as key management. 
The secure world has a small stack in order to reduce possible vulnerabilities, as it is intended to host trusted apps (\eg 210K LOC in a Linaro TEE, with 110K for the trusted OS and 100K for the secure monitor~\cite{cerdeira2020sok}). 

\para{TrustZone limitations for ML deployment.}
TrustZone has inherent limitations that make it incompatible with the practical implementation of ML services.
The first limitation is the \textit{reliance on the trusted OS} by the trusted apps.
Trusted apps' resources are controlled by the Trusted OS, which still has a large attack surface despite the smaller stack -- there have been a significant number of attacks that can compromise the Trusted OS~\cite{cerdeira2020sok}.
Prior work has suggested stronger isolation environments to address these issues~\cite{cerdeira2020sok, brasser2019sanctuary}.
The second limitation is \textit{memory}.
The memory size of the secure world is limited in current implementations (\eg 16$\sim$64MiB in OPTEE~\cite{mo2022sok}), which does not allow for the deployment of entire machine-learning models within the secure world.
%
%The memory size of the secure world is intentionally constrained (to reduce the size of the TCB)~\cite{mo2022sok} and fixed during runtime, which does not allow for the deployment of entire machine-learning models within the secure world. 
%
Finally, TrustZone has \textit{development cycle} limitations.
TrustZone was mainly designed to run trusted applications from platform-specific services (\eg from the original equipment manufacturers) rather than general developers~\cite{cca}.
In order to ensure that trusted apps do not contain vulnerabilities, vendors often place restrictions and security checks on developers who want to deploy trusted apps. 
Trusted apps also tend to be smaller in order to reduce the attack surface. 
This impedes the development of feature-rich apps. 
While prior work has attempted to get around these limitations~\cite{brasser2019sanctuary, bayerl2020offline, sun2022leap}, we look towards CCA as a possible solution as it is tailored towards secure deployment of general-purpose apps. 

\begin{figure}
    \centering
    \includegraphics[width=1\linewidth]{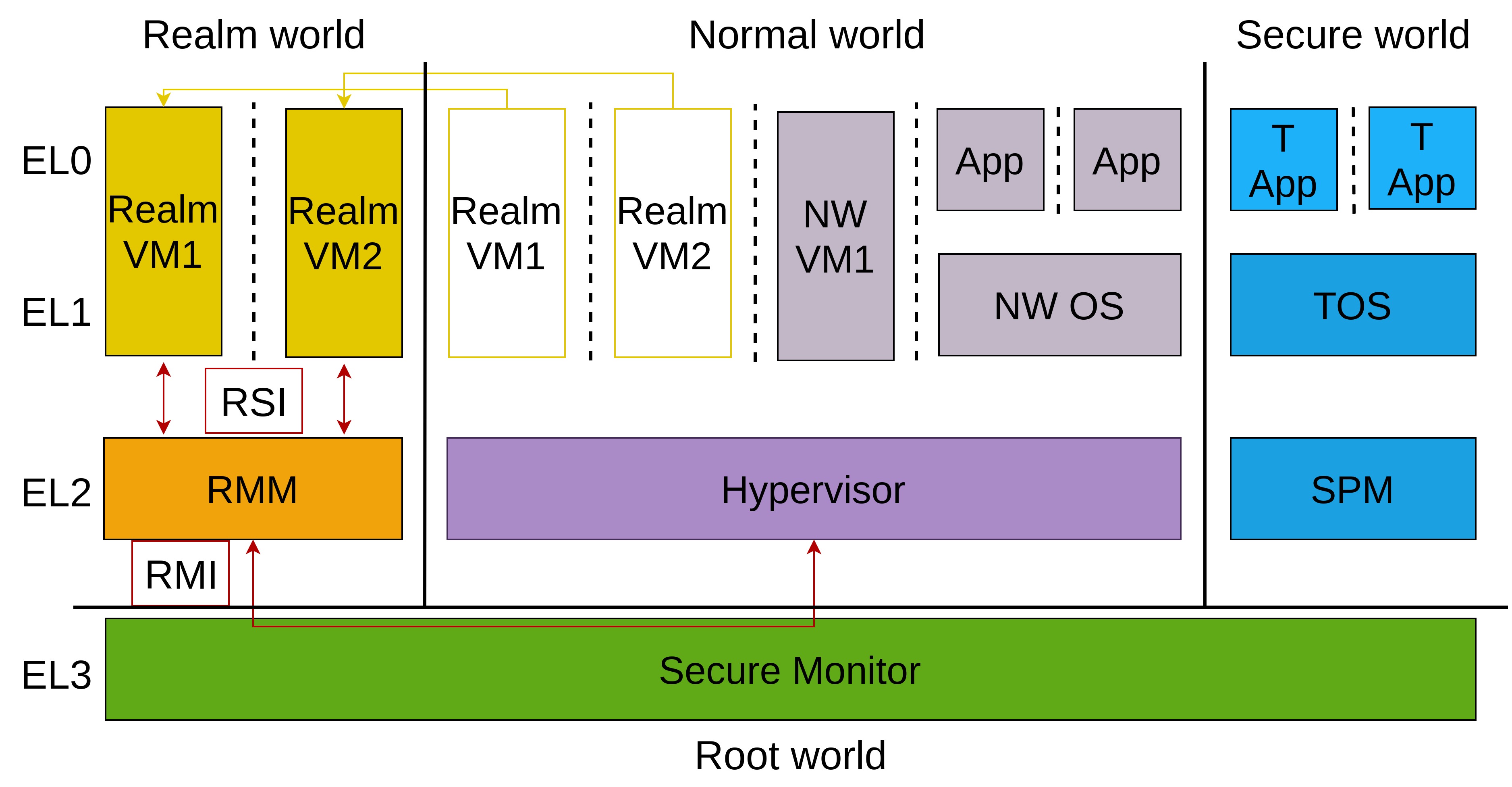}
    \caption{Arm CCA software architecture. CCA introduces two new execution environments: realm and root. CCA's architecture allows for the creation of dynamic, hardware-protected enclaves called \textit{realms}. Unlike TrustZone, the secure monitor runs in the root physical address (PA) space which is separate from the secure world PA.}
    \label{fig:Arm CCA software architecture}
\end{figure}

\para{Arm CCA.} Arm Confidential Computing Architecture (CCA) is a collection of hardware, software, and firmware extensions in the Armv9-A architecture~\cite{ccasite, cca, lienabling, li2022design}. 
As shown in Fig \ref{fig:Arm CCA software architecture}, in Arm CCA , there are two new execution environments (Realm and Root) in addition to the existing normal world and secure world execution environments. 
The root world is able to access all the other worlds.
The realm and the secure worlds cannot access each other's memory, \ie the realm world can access the realm and normal world memory while the secure world can access the secure and normal world memory.
As in TrustZone, the normal world cannot access the other worlds' memory.

The realm world architecture allows for the creation and execution of virtual machines (called realms). 
The hypervisor (in the normal world) performs initialization and memory allocation to the realms.
However, as the execution of a realm is isolated from the normal world, the hypervisor is not allowed to access it.
Thus, CCA introduces the Realm Management Monitor (RMM) -- a lightweight firmware in the realm world that manages communication between a realm and the hypervisor and ensures isolation between realms.
The communication interface between the hypervisor and the RMM is known as the 
the Realm Management Interface (RMI), whereas the interface between the realm and the RMM is known as the Realm Service Interface (RSI).
The RMI is used by the hypervisor to issue commands to the RMM for controlling the realm (\eg creation or termination of the realm). 
The RMM confirms the validity of these commands and then performs the requested action. 
The RSI is used by the realm to request services from the RMM (\eg create an attestation report). 

Transitions between worlds is managed by the Secure Monitor which resides in the root world.
We note that the root world has its own physical address space in CCA, unlike in TrustZone where the address space of the secure monitor was within the secure world.

\para{CCA for ML deployment.} 
CCA addresses the limitations of TrustZone for ML deployment, which makes it a viable candidate for \name.
First, CCA changes the trust relationship between a program and its supervisory software.
Unlike in TrustZone, where a compromised supervisory software (Trusted OS) can result in it gaining control of the trusted apps, CCA offers \textit{protection against a compromised hypervisor}.
While the hypervisor has the capability to manage realms and their resources, it is unable to access them, thereby ensuring confidentiality and integrity for the realms even when the hypervisor is compromised. 
Second, CCA offers \textit{flexible memory allocation} for the realms.
The memory that can be allocated to a realm is only limited by the system memory.
On realm termination, the hypervisor can reclaim the delegated memory and return it to the normal world. 
This allows for apps with large and dynamic memory requirements to be run with CCA.
Finally, CCA allows for \textit{flexible development}.
CCA is targeted towards general developers -- developers can easily deploy apps within realms without the need for business relationships with vendors. 

\para{Systems based on CCA.} 
As CCA is still under development, there is limited prior work in this space. 
Xu \etal introduced virtCCA~\cite{xu2023virtcca}, a virtualized CCA implementation on top of existing TrustZone hardware. virtCCA was meant to address the unavailability of hardware with CCA support. While CCA-compatible hardware has still not been released, we make use of Arm's released CCA-compatible software components to implement \name~\cite{fvp}. These components are actively being worked on and provide the necessary functionality required to deploy \name. 
Zhang \etal proposed Shelter~\cite{zhang2023shelter}, which provides user-space isolation in the normal world using CCA hardware primitives. Shelter is intended to be complementary to CCA's realms and was developed due to the nascent state of software development on CCA. In this work, we implement \name on CCA realms instead of opting for user-space enclaves in the normal world.
Sridhara \etal developed Acai~\cite{sridhara2024acai}, a system that allows CCA realms to securely access PCIe-based accelerators with strong isolation guarantees. Acai is complementary to our work; \name can be extended to use secure accelerators for running more complex models with mechanisms such as Acai.

\section{\name architecture}
In this section, we describe our system, \name, which uses CCA to run models in a private and attestable manner.

\subsection{System and Threat Model}

\para{System Model.} 
We consider three parties: \textit{model providers}, \textit{clients}, and \textit{a trusted verifier}.
A model provider is an entity that trains and deploys machine-learning models to clients.
However, before deploying an ML model, the provider needs to ensure that the client's suggested execution environment is trustworthy to store and use its model. 
The client is an end device (such as a smartphone or an IoT gateway) supporting the Armv9-A architecture.  
The client wants to use ML models for various purposes (\eg face recognition and speech processing) on their own device. 
The trusted verifier is an entity that provides verified realm images for clients to run.
A realm image contains all the software and dependencies required to deploy a realm within a client and run the provider's  model. 
The client obtains the verified image from the trusted verifier to deploy a realm within which the model is stored.

\para{Threat model.} 
We assume that the model provider does not trust the client -- the provider wants to ensure that the client only uses the model for inference, and wants to prevent the client from accessing any internals of the model. 
On the client side, we assume that the Secure Monitor and the RMM are trustworthy, and the normal world (including user-space apps and the hypervisor) are malicious. 
While CCA does not provide availability guarantees, we assume that the hypervisor, in addition to being unable to access a realm's content, does not interfere with the realm's creation and execution. 
We consider side-channel attacks out of scope for this paper.
Finally, we assume that the connection between the client and the model provider is encrypted and protected from external network adversaries and unintended decryption.

\subsection{\name Pipeline}

\begin{figure}[t]
    \centering
    \includegraphics[width=1\linewidth]{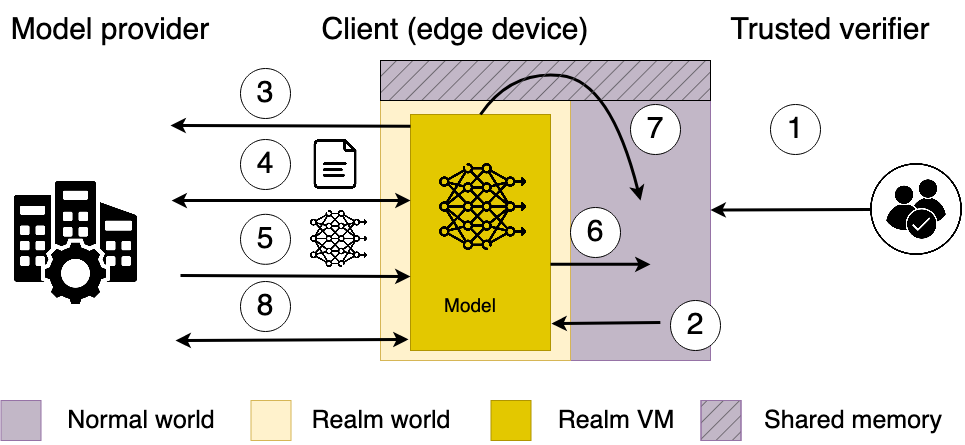}
    \caption{Overview of \name outlining the steps required for running a ML model on the client edge device. We show a simplified view of the normal and realm worlds within the client. The client's steps are (1) obtaining realm image from verifier (2) creating and activating a realm VM (3) establishing connection with provider (4) realm attestation (5) obtaining model from provider (6) announcing model readiness to normal world (7) running inference (8) performing model updates.}
    \label{fig:overview}
\end{figure}

We provide a description of the various steps involved in deploying a model on the client (Figure~\ref{fig:overview} shows an overview of \name). 

On the client, a normal world app starts the pipeline by obtaining a publicly available and verified realm image from the trusted third party (Step 1 in the figure).
The realm image contains a Linux operating system with all the software and dependencies required to run the provider's model. 

\para{Realm initialization.} The normal world app requests the hypervisor to initiate the realm creation process (Step 2).
The hypervisor delegates physical pages to the realm world, and requests the RMM to copy the content of the realm image to the delegated pages.
The RMM populates the realm pages and measures its contents. 
The hypervisor then sends the activation command to the RMM.
Once the realm is activated by the RMM, the hypervisor is able to give CPU execution to realm but it is no longer able to ask the RMM to populate new content. 

\para{Realm attestation.} 
Before sending a model to the client, the model provider needs to ensure that the model will be run within a realm that has been correctly initialized and is running without issues. 
This can be achieved by attestation of the realm. 
After booting, the realm initiates a TLS connection with the model provider (Step 3).
The model provider sends a request for an \textit{attestation report} to the realm, which forwards it to the RMM (Step 4).
The attestation report consists of two parts: attestation of the platform on which the realm runs (CCA token), and attestation of the state of the realm (realm token)~\cite{ccaattestation, sardar2023sok}.
The realm token consists of initial measurements taken during realm creation and runtime measurements. 
The RMM coordinates the attestation report creation: it obtains the CCA token from the root of trust, and the realm token from the realm, and assembles them into a report which it sends to the realm~\cite{rmmspec}. 
The realm, in turn, sends the report to the model provider. 
The model provider can verify the attestation report to decide the trustworthiness of the activated realm. 
On verification, it sends the model to the realm via the TLS connection (Step 5). 

\para{Inference.} After receipt of the model, the realm announces its ability to respond to inference queries to the normal world via the hypervisor (Step 6).  
The realm reads the input data from the normal world memory, feeds it into the model, obtains the inference, and writes the output to the normal world memory (Step 7).

\para{ML deployment life cycle.} In addition to the main operation of the model (\ie inference), \name needs to account for other aspects of a model's life cycle.
For example, the provider may want to impose a limit on the service it provides to the client -- this may be in the form of a time limit (or validity period) for which the model should be active or the number of inferences the model can provide.
Such functionality can be incorporated into the model-running code that is present in the realm image provided to the client.
Once the time/inference limit is reached, the realm can send a system call to the hypervisor asking for termination. 
On termination, the memory delegated to the realm is released back to the normal world.

Another scenario involved around model updates -- the realm may need to periodically query the provider to check for and receive updates (Step 8). 
The provider may need to verify the state of the realm before sending the updated model, which the realm can achieve by sending a runtime attestation report to the provider.

\subsection{Implementation}
As CCA is still under active development and a CCA-compatible board has not been released yet, we build a prototype of \name on Arm's Fixed Virtual Platforms (FVP)~\cite{fvp}. 
FVP are provided by Arm for early-stage development of software/firmware without the need for compatible hardware. 
We use the Armv-A Base Platform RevC version of FVP -- it supports the Armv8‐A architecture versions up to v8.7 and Armv9-A (which has the CCA extension)~\cite{fvp}.

\para{Software stack}. We use the latest release\footnote{AEMFVP-A-RME-2023.12.22} of the reference Arm CCA integration~\cite{CCAintegration} in our implementation. 
This integration consists of all the necessary source files and instructions to build binary files which are then provided to the FVP to start the simulation. 
The FVP boots the Secure Monitor, the RMM, and the hypervisor, respectively. 
The integration uses the Trusted Firmware-A~\cite{TF-A} and the Trusted Firmware implementation of RMM~\cite{RMM} as the Secure Monitor and RMM in the stack. 
It also uses linux-cca~\cite{linux-cca} as the hypervisor. 
There is no Trusted OS in the current version of the integration. 
As we do not run operations in the Secure world in our pipeline, we did not add Trusted OS to the stack.

The integration uses buildroot~\cite{buildroot} to create a customizable file system for the hypervisor as well as the realm. 
As we want to run a TensorFlow Lite model within the realm, we added C++ libraries to the default realm file system.
We then deploy the model within the realm. 

\name uses a folder mounted in the file system of the both realm and the hypervisor to exchange model inputs and outputs.
%\name uses shared memory between the realm and the normal world to exchange model inputs and outputs.
%
%We implement this shared memory as a folder mounted in the file system of both the realm and the hypervisor. 
%
We run applications in both the realm and the normal world. 
These applications check the contents of the shared folder.
On the realm side, the application checks whether there is input data available for inference.
If so, it reads the input data and feeds this into the model.
On the normal world side, the application checks for the presence of inference output from the realm.

We could not implement the attestation mechanism as FVP does not have the HES (Hardware Enforced Security) implemented, which is required for the attestation report (HES is required for the platform attestation token, which is a part of the attestation report)~\cite{ccaattestation}.
%
%We recommend that future versions of FVP provide this module to help implement attestation. 

\section{Preliminary evaluation}
In order to evaluate the overhead of running inference within a realm, we perform an evaluation between two scenarios for an image-recognition task.
In our baseline scenario, which does not provide confidentiality and integrity guarantees, the model and the code to run it are stored in a normal world virtual machine (VM). 
The model code reads the input data from the folder that is shared with a normal world app.
The code performs the inference and writes the output to shared memory.
In the second scenario, the model and code are stored within a realm VM. 
As in the baseline scenario, reading input data and writing inference output are performed via the shared folder with the normal world app. 

For our evaluation, we use a 16MB pre-trained TensorFlow Lite model MobileNet\_v1\_1.0\_224 \cite{Mobilenet}. This model is created by training a TensorFlow model on the ImageNet dataset at 224x224 resolution, and converting it to TensorFlow Lite.
We run our evaluation on a Lenovo ThinkCentre M75t Gen 2 with 16GB RAM and an 8-core AMD Ryzen 7 PRO 3700 processor (OS: Ubuntu 22.04.4 LTS).

\para{FVP considerations}.
FVP is instruction-accurate but not cycle-accurate, \ie it correctly reports the number of instructions required by an operation but not the time taken on real hardware.
As we cannot obtain accurate timing measurements without a CCA-compatible board, we rely on the number of instructions to obtain the overhead of running inference within the realm. 
Moreover, as we only have access to the total number of instructions executed on the FVP, we need a method to  obtain the number of instructions explicitly caused by a workload (\eg inference or realm creation). 
To report the real number, we first measure the number of instructions when the workload is not running and reduce that from the observed number of instructions when the workload is running on the FVP.
We also report timings for completeness, as the measurements may be useful for those intending to use FVP to experiment with CCA. 
\subsection{Inference overhead}
We create an instance of the realm and measure the total number of instructions for getting the inference result for 40 images.
We repeat this five times. 
Table~\ref{tab:evaluation} shows the average number of instructions for each inference process (writing one image to the shared memory by the normal world program, performing inference and writing output to the shared memory by the realm world program). 
As Table~\ref{tab:evaluation} shows, running the model inside the realm leads to $\times1.62$ overhead in the number of instructions executed for each inference process compared to running model inside the normal world VM. 
While timings are not accurate, we report them for completeness: on average, each inference takes $\approx$ 34.25 seconds in the realm VM, and $\approx$ 27.28 seconds in the normal world VM. 
The increase in the number of instructions is mainly due to the higher number context switches the hypervisor needs to manage the realm and handle its interrupts. 
While the hypervisor has direct access to the normal world VM, its access to the realm first has to go through the Secure Monitor and then to the RMM, resulting in more context switches and hence, more instructions. 

\begin{table}[t!]
  \centering
\caption{Mean (standard deviation) number of instructions for each inference over five experiment runs.}
    \resizebox*{\linewidth}{!}{%
 % \begin{tabular}[c]{ p{0.4\linewidth} p{0.5\linewidth}}
  \begin{tabular}[c]{ l  c}
    \toprule
    \textbf{Scenario}   & \textbf{Number of Instructions (Millions)}\\
    \midrule
    Normal world VM  &  222.2 (46.5) \\
     Realm VM  &  361.6 (4.4) \\
    \bottomrule
  \end{tabular}
    }
  \label{tab:evaluation}
\end{table}

\subsection{Realm setup}

We also report the instructions and the time required to create and terminate a realm VM as well as a normal world VM when the size of the image is 98MB (55MB for the file system and 43MB for the kernel)\footnote{We allocate 300 MB RAM to create the VM, which is sufficient for the VM to function correctly without being too large.}.
We repeat each observation five times.
As shown in Table~\ref{tab:evaluationsetup}, creating a realm VM creates a significant overhead ($\times26.62$) as compared to the normal world VM. 
The increased overhead is $\times9.23$ for termination. 
On average, realm creation and termination take 6:21 min and 1:01 min, whereas the normal world VM creation and termination take 1:31 min and 0:08 min respectively.  
We observe that the size of the initial content to be populated intro the realm has a significant effect on the boot time.
For instance, when we increase the size of the realm image to 139MB (96MB for the file system and 43MB for the kernel)\footnote{We allocate 400MB RAM in this scenario.}, we observe that, on average (over five iterations), it takes $\approx$ 27,190 million instructions to create the realm VM and $\approx$ 725 million instructions to create the normal world VM. 
The overhead is around $\times37.50$, which is a significant increase compared to the previous comparison.
Thus, we recommend that the realm image only contain code required to implement necessary functionality.  
Reducing realm image size would also facilitate multiple realms running in parallel. 

\begin{table}[t!]
  \centering
\caption{Mean (standard deviation) number of instructions for booting and termination over five experiment runs. Image size is 98MB.}
    \resizebox*{\linewidth}{!}{%
  \begin{tabular}[c]{ l  c}
    \toprule
    \textbf{Scenario} & \textbf{Number of instructions (Millions)}\\
    \midrule
    Realm VM boot              & 18,880.6 (1,655.3) \\
    Normal VM boot                   &  709.8  (6.7) \\
    Realm VM termination         &  970.0 (98.9) \\ 
    Normal VM termination                   &  105.1  (0.2) \\
    \bottomrule
  \end{tabular}
    }
  \label{tab:evaluationsetup}
\end{table}

\section{Considerations for ML deployment using CCA}
Our prototype of \name enables us to take initial steps towards using CCA to run ML models. 
However, there are several other factors to take into account for deploying the entire ML cycle, which may require modifications to the CCA architecture.

First, we only consider a scenario where adversaries would not have access to the model.
However, there are attacks on the data pipeline, as the inputs and outputs to the model are not protected~\cite{mo2022sok}. 
For example, adversaries can poison the input data to the model, or run attacks on the ML inferences, as these are not protected in our current implementation of \name (input data and inferences are stored in the normal world). 
The CCA architecture also does not currently offer dedicated solutions to secure the inputs and outputs to the model. 
Potential solutions include exploring how secure peripherals can be used with realms~\cite{schneider2022sok, sridhara2024acai} and integrating detection mechanisms within the realm. 

We could not implement the exact attestation mechanism as described in the CCA specification, as FVP does not currently have the HES (Hardware Enforces Security) module to create the platform attestation token. 
As attestation is an important step in the ML deployment pipeline, we recommend that future versions of FVP provide this module to help implement and test attestation. 

We also envision that models from multiple (possibly, competing) providers may concurrently run on a user's device.
Measuring performance when running multiple realms in parallel, and extending CCA support for sharing of resources among realms (\eg access to peripherals) are necessary to realize this scenario. 

Furthermore, we need to employ mechanisms to ensure that the ML pipeline is operational in accordance with the agreement between the client and the provider, especially in the absence of a trusted verifier \eg if the model provider supplies the realm image.
The client needs assurance that the model provider's code will not use its data for purposes outside the agreed-upon ones and the model provider needs assurance that the client uses its services only for the time or inference limit it has agreed to. 
These mechanisms have an impact on all aspects of the pipeline, from model inference to updates to termination.
Extending the CCA architecture to include functionality for policy enforcement and improved runtime realm attestation would assist in these changes. 

Finally, CCA does not provide availability guarantees -- the hypervisor controls the creation and execution of realms. 
The hypervisor can, thus, cause denial of service for the client, \eg by preventing creation of the realm for model deployment. 
Exploring methods to provide greater availability guarantees would be interesting future work.

\section{Conclusion}
In this work, we introduced \name, a framework that uses CCA to provide ML model protection and attestation on edge devices.
We used the FVP simulator to develop and evaluate an initial version of ML inference.
Our results indicate that it is feasible to use FVP to build CCA-compatible prototypes of various stages of the ML deployment cycle.
We also find that running a model within a realm for inference incurs an overhead of 1.7 times the number of instructions required for running it in a normal world virtual machine. 
At the same time, we discover various challenges that may require changes, not only to the FVP platform, but also to the underlying CCA architecture to fully realize the vision of private and attestable learning at the edge.  

\begin{acks}
We thank Jon Crowcroft for his invaluable suggestions that helped improve the paper.
This work was funded by the EPSRC Open Plus Fellowship (EP/W005271/1) and an Amazon Research Award.
\end{acks}
% \begin{acks}
% \end{acks}

\bibliographystyle{ACM-Reference-Format}
\bibliography{bibliography}

% \appendix
% \input{parts/appendix}

% \clearpage
% \onecolumn

% % \MarkupsHowto % prints markups help page

\end{document}